\documentclass[12pt]{article}
\usepackage{graphicx}
\usepackage{amsmath}

\usepackage{arxiv}

\usepackage[utf8x]{inputenc}

\title{ An issue of determination of the centrality in nucleus-nucleus collisions}
\author{R.M. Djilkibaev\\ 
     \normalsize   Institute for Nuclear Research, 
     Moscow, Russia 117312 \\ \normalsize  email: {rmd@inr.ru}}

\begin{document}

\maketitle

\begin{abstract}

The accuracy of measuring the total energy of the spectators in the
FHCal calorimeter, depending on the fluctuations in the number of nucleons 
and the hadron shower is obtained.
It is shown that, from the point of view of statistics, 
the process of registering photoelectrons in a photomultiplier (PM) is similar 
to the process of registering spectator nucleons in a calorimeter.
The measurements of single photoelectrons in the PM are in 
good agreement with the obtained formula for the relative 
variance of the charge detected by the PM.

\end{abstract}

The purpose of this work is to evaluate the accuracy of determining the centrality
by measurement
of the total energy E of spectator nucleons in the FHCal calorimeter,
taking into account fluctuations
of the number $ N_s $ of spectator nucleons and fluctuations
of energy release during registration of a hadron shower in a calorimeter.
Nucleus-nucleus collision calorimeter
registers in each event simultaneously a different number of $ N_s $ spectator nucleons
(see Fig. 1). 
In this case, the kinetic energy of nucleons with good accuracy is equal to
$ E_b $ beam energy. 
Such conditions cannot be created for carrying out calibration
measurements of the FHCal calorimeter, close to the experiment.
From the point of view of statistics, there is a process similar
to the process of registering spectator nucleons in a calorimeter, this
is the process of registering photoelectrons in a photomultiplier (PM).
Calibration measurements are possible for this process.
A short flash of light with a duration of $\simeq$ 50 ns leads to
the instantaneous birth
of a different number of $N_{pe}$ photoelectrons on the photocathode 
of the PM, in each event.
In this case, each photoelectron forms
an independent avalanche process in the dinode system of the PM,
similar to the formation of showers in the FHCal calorimeter.
Paper \cite{gol} gives an estimate
of the accuracy of determining the centrality parameter
$\sigma_{b}/b \simeq \sigma_{E}/E \simeq \sigma_{N_s}/N_s$
by measurement of
the total energy E of spectator nucleons in the FHCal calorimeter.

\begin{figure}[htb!]
  \centering
  \includegraphics[width=0.8\textwidth]{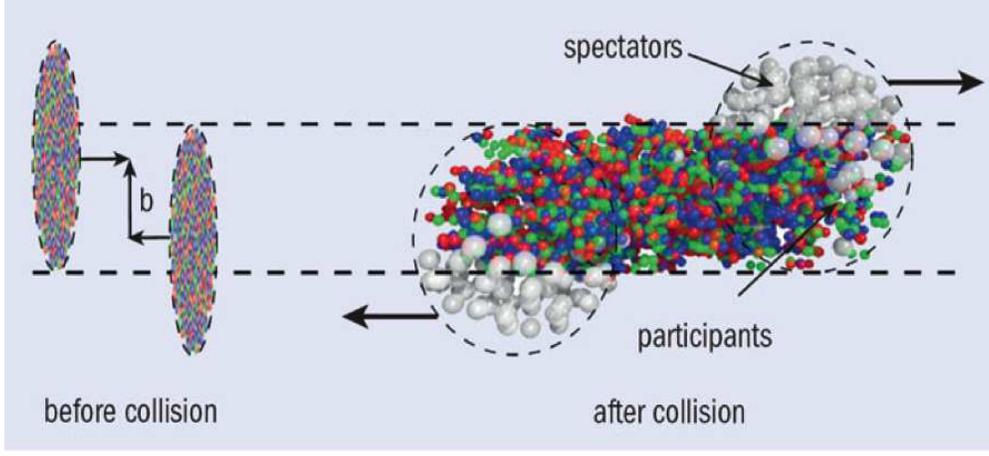} 
  \caption[Short caption.]{
   Schematic picture of a nucleus-nucleus collision with the impact parameter b.
}
\label{fig:setup}
\end{figure}

The total energy release E in the calorimeter is expressed as the sum
of the random energy releases of the spectator nucleons in the calorimeter
$E = \sum\limits_{i=1}^{N_s}{E_{s}^{i}}$.
This process is stochastic, in which the same energy release can be obtained for a different
number of spectator $N_s$ nucleons.
The total energy release E in the calorimeter
can be represented as the product of two
random variables $N_s$ and $\overline{E}_s$ - the sample average energy release,
in the form:

\begin{equation}
E = \sum\limits_{i=1}^{N_s}{E_{s}^{i}} = N_s \cdot \overline{E}_s; ~~~~~ 
\overline{E}_s = \frac{1}{N_s} \sum\limits_{i=1}^{N_s}{E_{s}^{i}}
\label{eqP}
\end{equation}

The relative error of the values $x = u \cdot v$ and $y = u/v$,
depending on two (u, v) independent random variables,
can be expressed as \cite{bev}:

\begin{equation}
\frac{\sigma_x^2}{x^2} = \frac{\sigma_u^2}{u^2} + \frac{\sigma_v^2}{v^2}; 
~~~~~~~~\frac{\sigma_y^2}{y^2} = \frac{\sigma_u^2}{u^2} + \frac{\sigma_v^2}{v^2} 
\label{eqP0}
\end{equation}

In our case, the relative error $\sigma_{E}/E$
is expressed as follows:

\begin{equation}
\sigma_{E}^2/E^2 = \sigma_{N_s}^2/N_s^2 + \sigma_{\overline{E}_s}^2/\overline{E}_{s}^2 
\label{eqP1}
\end{equation}

Relative variance of the sample average energy release $\overline{E}_s$ 
decreases with increasing $N_s$ compared to the relative variance 
of a single shower $E_s$ as follows \cite{bev}:

\begin{equation}
\sigma_{\overline{E}_s}^2/\overline{E}_{s}^2  = \frac{1}{N_s} \sigma_{E_s}^2/E_{s}^2
\label{eqP2}
\end{equation}

The error $\sigma_{N_s}$ is determined by the Poisson statistics 
and is equal to $\sigma_{N_s}^2 = N_s$.
Relative error of measurement of the nucleon energy in
the FHCal calorimeter \cite{gol} is $ \sigma_{E_s}^2/E_s^2 = a^2/E_s$, 
where the value a = 0.56.
As a result, the expression for the relative error $\sigma_{E}/E$,
can be written as follows:

\begin{equation}
 \sigma_{E}^2/E^2
 = \frac{1}{N_s} ( 1 + \frac{a^2}{E_s} )
\label{eqP4}
\end{equation}

In the experiment, it is necessary to determine
the number of $ N_s $ spectator nucleons in each event.
The total energy release in the calorimeter $E^{'}$ can,
in the first approximation, be expressed in terms of the estimated number of$ N_s^{'} $ spectator nucleons
in the form
$E^{'} = N_s^{'} \cdot E_b$, given that the kinetic energy of the nucleons with good accuracy is
$ E_b $ of the beam energy.
Then the relative variance of the number of $N_s$ spectator nucleons can be expressed in terms of the total
energy $E^{'}$ and the sample average energy release
of the nucleon in the calorimeter $\overline{E}_s^{'}$
based on the following expression:

\begin{equation}
N_s = E^{'}/{\overline{E}_s^{'}}; ~~~~~ 
\overline{E}_s^{'} = \frac{1}{N_s^{'}} \sum\limits_{i=1}^{N_s^{'}}{E_{s}^{i}}
\label{eqP}
\end{equation}

In our case, the relative error $\sigma_{N_s}/N_s$
is expressed similarly to (2) as follows:

\begin{equation}
\sigma_{N_s}^2/N_s^{2} = \sigma_{E^{'}}^2/E^{'2} + \sigma_{\overline{E}_s^{'}}^2/\overline{E}_{s}^{'2} 
\label{eqP1}
\end{equation}

As a result, the expression for the relative error $\sigma_{N_s}/N_s$, given the expression (4),
can be written as follows:

\begin{equation}
 \sigma_{N_s}^2/N_s^{2}
 = \frac{1}{N_s^{'}} ( 1 + \frac{a^2}{E_s} )
\label{eqP4}
\end{equation}

In work \cite{kur}
for the relative variance of the total energy E and the number of $N_s$ 
spectator nucleons, the following expressions are obtained:

\begin{equation}
\sigma_{E}^2/E^{2} = \sigma_{N_s}^2/N_s^{2}  = \frac{1}{N_s^{'}} + a^2/E_s 
\label{eqP5}
\end{equation}

The unexpected conclusion follows from this formula
that for large $N_s^ {'}$,
large energy release in the calorimeter,
relative measurement errors
$\sigma_{N_s}/N_s $ and $\sigma_{E}/E $
do not depend on the estimated $N_s^{'}$ and are determined by the relative error of measuring
the energy of a single nucleon.

This is a consequence of the fact that the derivation of formula (9) in \cite{kur}
is based on an incorrect
notation of the stochastic process,
where the measured energy of a single nucleon $E_s$ is taken, instead of the  
$\overline{E}_s^{'}$ sample average energy release.

The process of registration of single photo-electrons of the PM is similar,
from the point of view of statistics, to the process of registration 
of spectator nucleons in a calorimeter.
Photons form photoelectrons in the PM photocathode with a probability equal
to the quantum efficiency of the photocathode $\epsilon \simeq 20\%$.
After the formation of the photo-electron, an avalanche-shaped charge gain 
occurs in the dinode system of the PM.
The process of formation of $N_{pe}$ photo-electrons is stochastic.
The amplification process in the PM is also stochastic.
Each photoelectron is amplified in a PM with a different gain.
The registered charge Q consists of the sum of$N_{pe}$ of random amplification processes.
The total charge Q in a single event can be expressed in simplified form
as the product of two random variables $N_{pe}$ and the average gain $G_{avr}$ of the PM
as follows:

\begin{equation}
Q = e \sum\limits_{i=1}^{N_{pe}}{G_i} = e \cdot N_{pe} \cdot G_{avr}; ~~~~~ 
G_{avr} = \frac{1}{N_s} \sum\limits_{i=1}^{N_{pe}}{G_{i}}
\label{eqP}
\end{equation}

where e is the charge of the electron, $Gi_i$ is the gain for the $i$ - th photo-electron.
The relative fluctuation of the detected charge depends on the fluctuation of the number
of photo-electrons and the fluctuation of the PM gain when registering a single
photoelectron $\sigma_{G_i}$, as follows \cite{bir, dor}:


\begin{equation}
\sigma_{Q}^2/Q^2 = \sigma_{N_{pe}}^2/N_{pe}^2 + \sigma_{G_{avr}}^2/G_{avr}^2 = 
\frac{1}{N_{pe}} ( 1 + \frac{\sigma_{1}^2}{Q_1^2} )
\label{eqP1}
\end{equation}

where $\sigma_1$ is the variance, $Q_1$ is the average charge for a single photo-electron.

The process of formation of the average number of $\mu$ photoelectrons from the light flash
of an LED is stochastic and is described by Poisson statistics.
The probability of registering n photoelectrons is $ P(n) = \mu^n e^{-\mu}/n!$.
The charge registered by the PM is composed of the sum of random 
amplification processes.
In paper \cite{bel}, a method for calibrating the PM by measuring 
single photoelectrons is proposed.
The spectrum of charges $Q(x)$ registered by the PM is described as the sum of signals
from background processes $B(x)$ and signals from photo-electrons $P(x)$
as follows $Q (x) = N_0(B (x) + P (x))$ \cite{bel}:

\begin{equation}
\begin{aligned}
B(x) = \Bigl \lbrace \frac{(1- w)}{\sigma_0 \sqrt{2\pi}}e^{-\frac{(x- Q_0)^2}{2\sigma_0^2}} +
w\cdot \theta(x - Q_0)\cdot \alpha \cdot exp(-\alpha (x - Q_0))
\Bigr \rbrace \cdot e^{-\mu} \\
P(x) = \sum\limits_{n=1}^{\infty}\frac{\mu^n e^{-\mu}}{n!} \frac{1}{\sigma_1 \sqrt{2n\pi}}
exp(-\frac{(x - Q_0 - Q_{sh} - n\cdot Q_1)^2}{2n\sigma_1^2}),  Q_{sh} = w/\alpha 
\end{aligned}
\label{eqP1}
\end{equation}

where $Q_0$ and $ \sigma_0$ are the mean and variance of the pedestal when measuring the detected
charge of the PM in the absence of primary photo-electrons (n = 0),
the parameters w and $\alpha$ describe the contribution of background processes, $\theta(x)$ - 
the step function, $Q_1$ and $ \sigma_1$ are the mean and variance of the signal from a single photoelectron,
$Q_{sh}$ - offset of the pedestal associated with the background,
$N_0$ – the normalized number of events.

The total number of parameters describing the function $Q(x)$ is eight.
When registering a large number of photoelectrons $\mu \gg 1$, 
the Poisson statistics transfer to  the Gaussian statistics and the 
average charge and the charge variance  from the light signal will be equal to
$\mu Q_1$ and $\sqrt{\mu(\sigma_1^2\ + Q_1^2)}$, respectively.
It follows that the relative charge variance for large $\mu$ is equal to:


\begin{equation}
\sigma_{Q}^2/Q^2 = 
\frac{1}{\mu} ( 1 + \frac{\sigma_{1}^2}{Q_1^2} )
\label{eqP1}
\end{equation}

The scheme for measuring single photoelectrons using a fast LED and
a pulse generator is shown in Figure 2a.
Figure 2b shows a typical PM signal from a single photoelectron from the output of the amplifier and
the generator signal for starting the LED.
The LED was started from a rectangular signal with an amplitude of 2.95 and 3.25 volts and
a duration of 60 seconds from the generator.
With the help of a diaphragm, the attenuation of the light flux was selected for
measuring single photoelectrons.
Measurements of the pulse shape of single photoelectrons of the PM were carried out using
12-bit 16-channel ADC converter CAEN DT5742 (WFD-waveform digitizer) \cite{cae}.
The device operated at a frequency of 2.5 GHz with the measurement of the signal 
amplitude at 1024 points with an interval of 0.4 seconds.

\begin{figure}[htb!]
  \centering{\hbox{
 \includegraphics[width=0.5\textwidth]{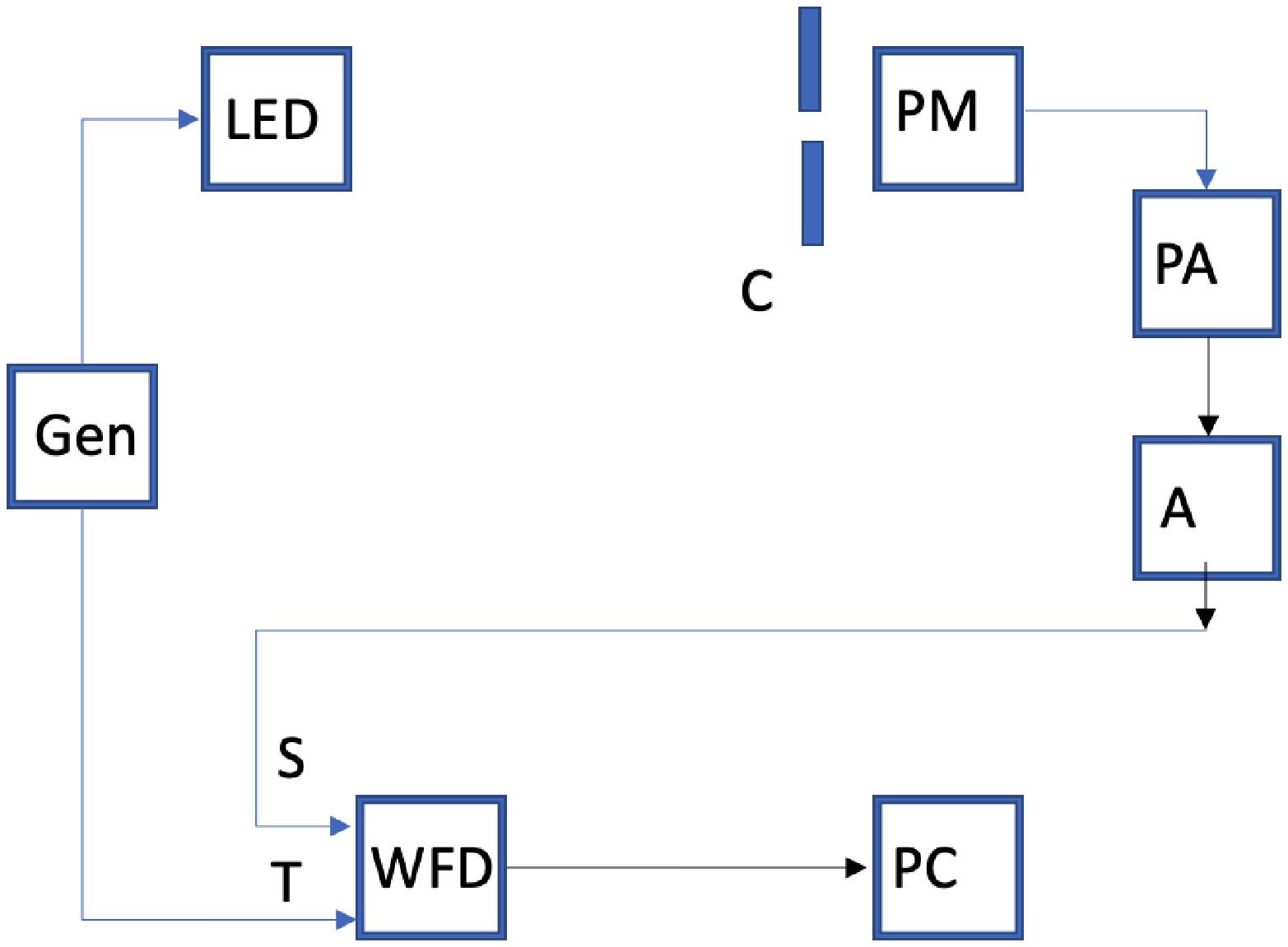}
 \includegraphics[width=0.5\textwidth]{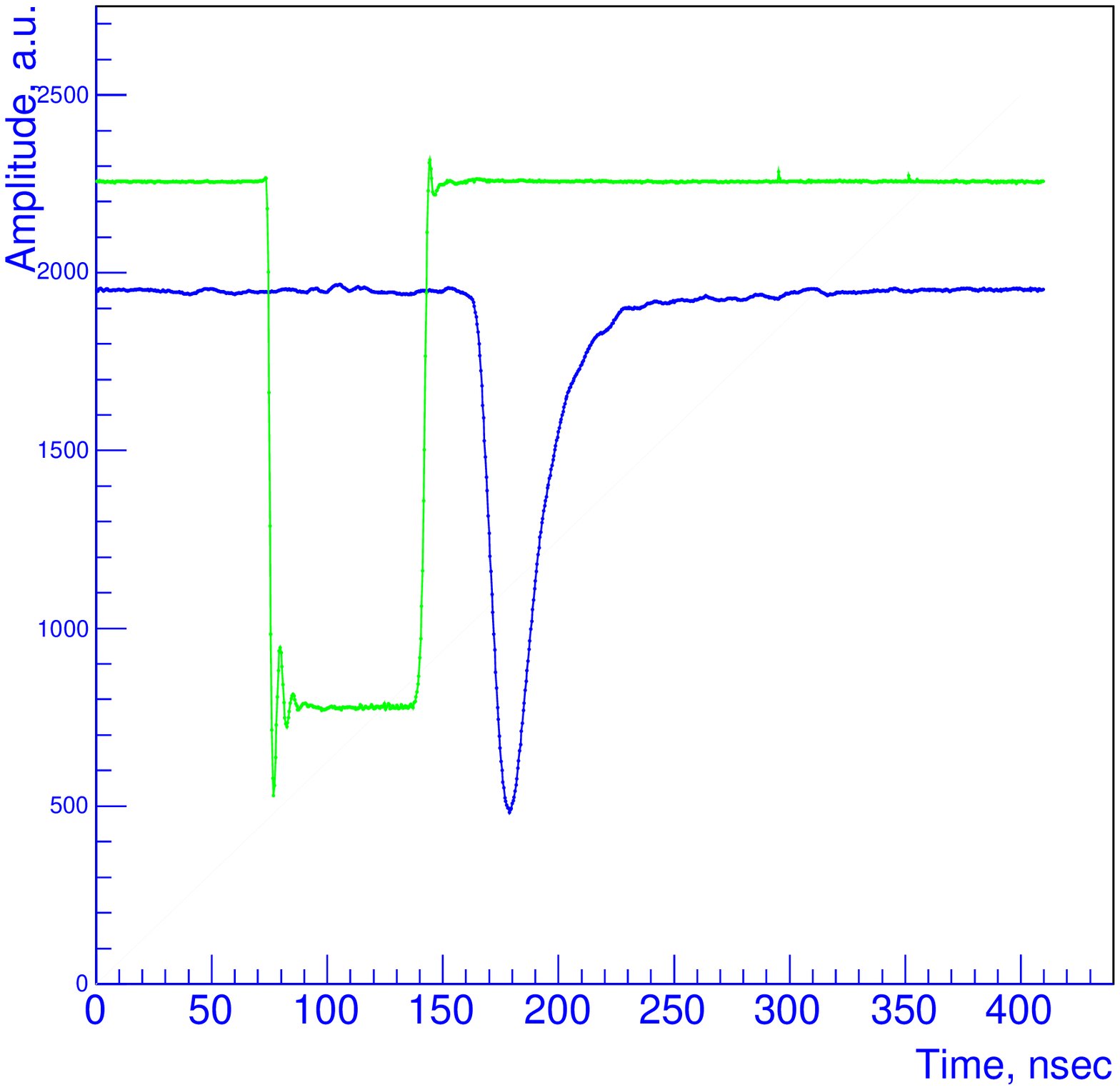}
 }}
  \caption{
(a) Block scheme for measuring single photoelectrons using a fast LED,
where Gen is the generator, C is the diaphragm, PM is the XP2020, PA is the preamp,
A - amplifier, WFD - ADC CAEN DT5742, PC - computer;
(b) - a typical PM signal from a single photoelectron from the output of the amplifier and
a generator signal to start the LED.
}
\label{fig:DiffSpec}
\end{figure}

The results of the fit of the measured charge from single photoelectrons with the amplitudes
of the LED start signal equal to 2.95 and 3.25 V and a duration of 60 nsec
are shown in Figures 3a and 3b, respectively.
The components of the spectrum of the detected charges $Q(x)$: the background and contributions from
one, two, three, etc. photoelectrons are also shown in Figure 3.


\begin{figure}[htb!]
  \centering{\hbox{
 \includegraphics[width=0.5\textwidth]{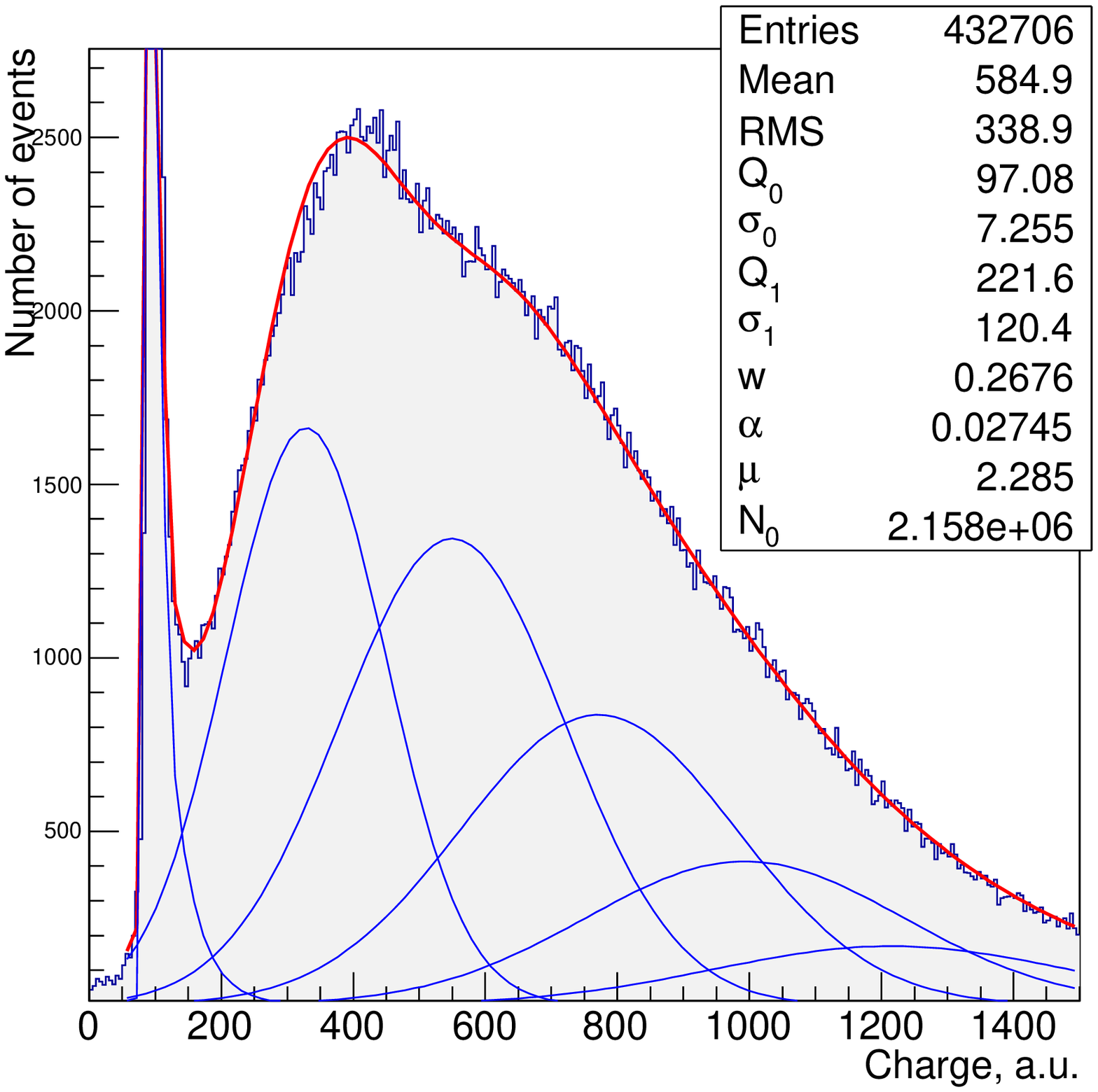}
 \includegraphics[width=0.5\textwidth]{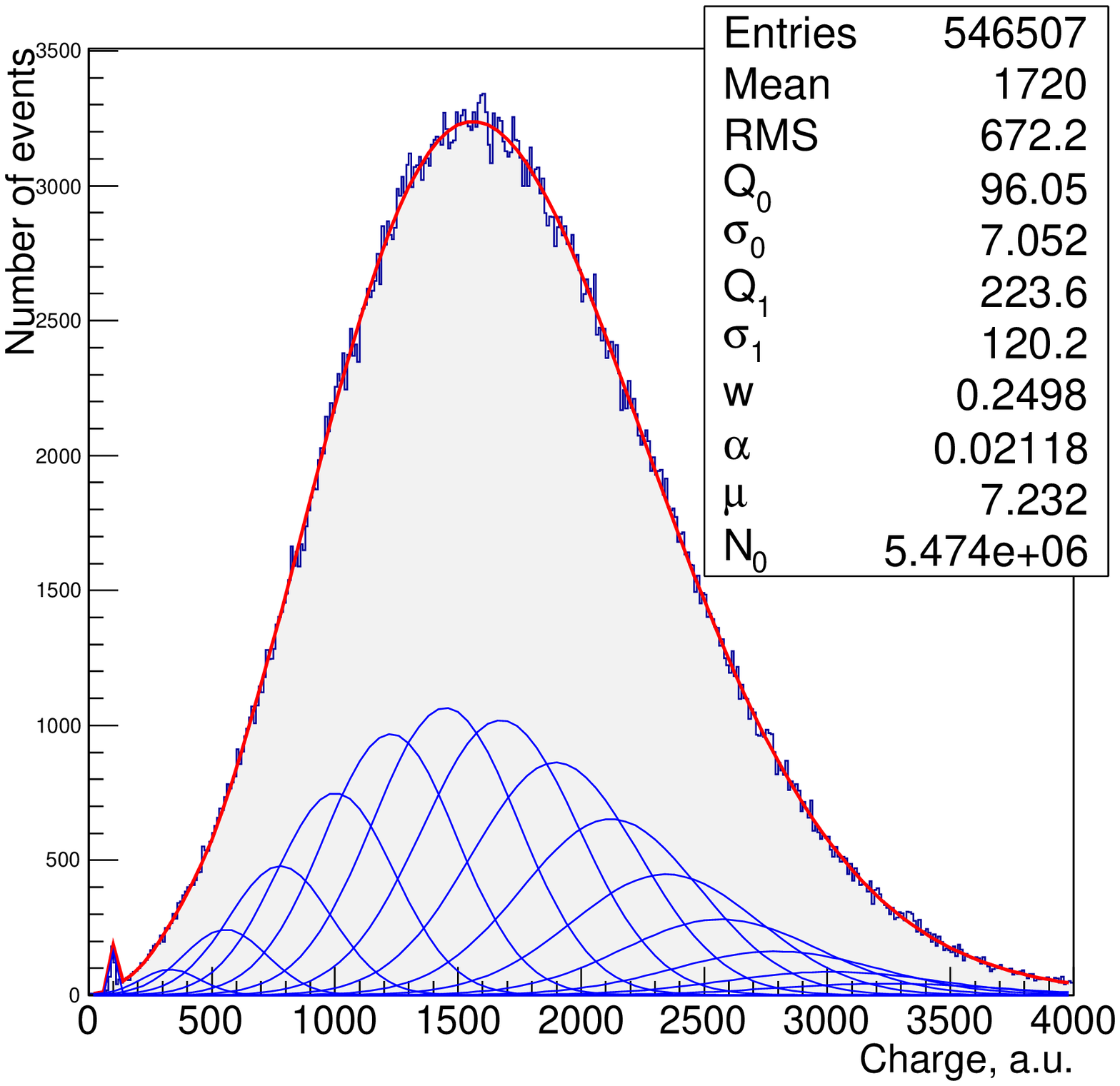}
 }}
  \caption{
(a) - spectrum of the detected charge of the PM and the result of the fit of the spectrum
distribution Q(x) for the average number of photoelectrons $\mu $= 2.28;
(b) - spectrum of the detected charge of the PM and the result of the fit for the average number
photoelectrons $\mu$= 7.23.
Also shown are the components of the spectrum: the background and the 
contributions from one, two, three, etc.  photoelectrons.
}
\label{fig:DiffSpec}
\end{figure}

A comparison of the measured charge spectra (Figs. 3a and 3b) from single photoelectrons
demonstrates the statement about the transition of the Poisson statistics 
to the Gaussian statistics with
an increase in the average number of photoelectrons registered by the PM  from a light flash.
An estimate of the relative variance of the measured charge from the distribution shown in
Figure 3b, with an average of $\mu = 7.2$, gives a value equal to $\sigma_Q/Q = 0.43$.
The estimate of the relative charge variance by formula (13) at $\mu = 7.2$
gives $\sigma_Q/Q = 0.42$, which is in good agreement with the experiment.

Paper \cite{bel} presents the results of charge measurements for the average number
of photoelectrons equal to $ \mu = 6.7$.
From this spectrum, we can estimate
the relative charge variance, which is $\sigma_Q/Q = 0.42$.
The relative variance of the charge calculated by formula (13) 
is $ \sigma_Q/Q = 0.43$.
This also agrees well with the experiment.

Estimation of the value of the relative variance of the charge by formula (9)
from \cite{kur} at $\mu = 7.2$, gives $\sigma_Q/Q = 0.65$ and 
does not agree ($48 \%$) with the measured value of 0.43.
With an increase in the average number of photoelectrons registered by the PM, this difference will only
increase.

\section*{Conclusions}  

The accuracy of measuring the total energy of the spectators in the
FHCal calorimeter, depending on the fluctuations in the number of nucleons 
and the hadron shower is obtained.
It is shown that, from the point of view of statistics, 
the process of registering photoelectrons in a photomultiplier (PM) is similar 
to the process of registering spectator nucleons in a calorimeter.
The measurements of single photoelectrons in the PM agree well with
the obtained formula for the relative variance of the charge detected by the PM,
and do not agree with the formula obtained in \cite{kur}.

\section*{Acknowledgments}  

In conclusion, I would like to express my gratitude for the useful 
discussions and comments to 
A. Ivashkin, F. Guber, A. Kurepin, I. Tkachev.
The work was carried out using the equipment of the collective
user center "Accelerator Center for Neutron Studies of the Structure
of Matter and Nuclear Medicine of the INR RAS" with the support of the Ministry
of Education and Science of the Russian Federation under 
Grant Agreement No. 14.621.21.0014 of 28.08.2017,
unique identifier RFMEFI62117X0014.



\begin{thebibliography}{99}
\bibitem{gol} M.B. Golubeva et al. Forward Hadron Calorimeter (FHCal). 
Technical Design Report for the MPD Experiment, Tech. Rep., JINR, Dubna, 2017, 
URL http://mpd.jinr.ru/wp-content/uploads/2017/08/MPD-TDR-FHCal-v9-1.pdf.
\bibitem{bev} Philip  Bevington, D. Keith Robinson; 
Data reduction and error analysis for the physical sciences, McGraw-Hill, Inc 2003
\bibitem{kur} A.B. Kurepin,A.G. Litvinenko, E.I. Litvinenko https://arxiv.org/abs/1901.06508,
Physics of Atomic Nuclei, 2020, Vol. 83, No. 9, pp. 1359-1362.
\bibitem{bir} J.B. Birks, The Theory and Practice of Scintiilation Counters, New York, Pegamon, 1967.
\bibitem{dor} P. Dorenbos, J.T.M. Hass, C.W.E. Eijk, IEEE Trans. Nucl. Scie. 1995. V. 42. No 6. P. 2190. 
\bibitem{bel} E.H. Bellamy, G. Bellettiny, J. Budagov  et al., Nucl. Instruments and Methods in Physics Research,  
1994, v. A339, p. 468-476
\bibitem{cae} CAEN DT5742, 16 channel 12 bit waveform digitizer, http://www.caen.it 
\end{thebibliography}
\end{document}